\documentclass[journal,twocolumn,xtable=color]{IEEEtran}
\usepackage{ifpdf}
\usepackage{cite}
\usepackage[pdftex]{graphicx}
\usepackage{amsmath}
\usepackage{algorithmic}
\usepackage{array}
\usepackage{tabularx}
\usepackage{booktabs}
\usepackage{tabu,multirow}
\usepackage[pdftex,pdfpagelabels,bookmarks,hyperindex,hyperfigures]{hyperref}
\usepackage[open]{bookmark}
\usepackage{subcaption}
\usepackage{makecell}
\usepackage{adjustbox,lipsum}
\usepackage{float}
\usepackage{afterpage}
\usepackage{soul}
\usepackage{xcolor}
\usepackage{chngcntr}
\usepackage{comment}
\newcommand{\cmmnt}[1]{\ignorespaces}
    \makeatletter
    \newcommand{\thickhline}{%
        \noalign {\ifnum 0=`}\fi \hrule height 1pt
        \futurelet \reserved@a \@xhline
    }
    \newcolumntype{"}{@{\vrule width 1pt}}
    \makeatother
\newcolumntype{?}[1]{!{\vrule width #1}}
\begin{document}
\title{Towards a Hybrid RF/Optical Lunar Communication System (LunarComm)}
\author{
    \IEEEauthorblockN{Waseem~Raza\IEEEmembership{}, Ethan~Abele\IEEEmembership{}, John~O'Hara,~\IEEEmembership{Senior Member,~IEEE}, Behnaz~Sadr\IEEEmembership{}, Peter~LoPresti,~\IEEEmembership{Member, IEEE}, Ali~Imran, \IEEEmembership{Senior Member,~IEEE},~Wooyeol~Choi, \IEEEmembership{Senior~Member,~IEEE},~Ickhyun~Song\IEEEmembership{}, Serhat~Altunc\IEEEmembership{}, Obadiah~Kegege\IEEEmembership{}, and Sabit~Ekin*,~\IEEEmembership{Senior Member,~IEEE}}\\

\thanks{\textit{*Corresponding author: Sabit Ekin}.}

\thanks{W.~Raza and A.~Imran are with the University of Oklahoma; E.~Abele, J.~O'Hara, W.~Choi, and S.~Ekin  are with Oklahoma State University; B.~Sadr and P. LoPresti are with University of Tulsa; I.~Song is with the Hanyang University, Seoul, South Korea; S. Altunc and O.~Kegege are with NASA Goddard Space Flight Center.}

}

\vspace{-2cm}
\maketitle
\begin{abstract}
The prospect of mankind returning to the Moon has garnered a great amount of attention in recent years. Dozens of lunar missions are planned for the coming decade which will require the development of a sustainable communication infrastructure with high data rates and minimal latency. Space communication systems thus far have relied on Radio Frequency (RF) links alone, but recent developments in laser communications have demonstrated that Free Space Optical (FSO) links can achieve much higher data rates. Upon considering the respective benefits and drawbacks of RF and FSO links, we make a case for the integration of these two technologies into a hybrid RF/FSO lunar communications architecture which leverages small satellites in a Low Earth Orbit (LEO) constellation. We include a case study for this technology designed in Analytical Graphics' Systems Tool Kit (STK) software. Results are presented in terms of chain access duration, propagation delay, transmission loss, Signal-to-Noise Ratio (SNR), and Bit Error Rate (BER). This architecture shows potential to revolutionize extraterrestrial communications and pave the way for highly ambitious future missions in space.
\end{abstract}
\begin{IEEEkeywords}
RF, FSO, Hybrid, Lunar Communications, LunarComm, Lunar Networking, LunaNet, Laser, Optical, Space, Network, Moon, LEO, SmallSat, STK.
\end{IEEEkeywords}
\section{Introduction}
\IEEEPARstart{T}{HE} National Aeronautics and Space Administration (NASA) has announced the Artemis project which aims to send the first woman and next man to the Moon by $2024$ and establish a permanent presence on the Lunar surface~\cite{2020_FSOC_CubeSats_LLO}. There are many reasons for this renewed interest in lunar exploration including that the Moon can be the basis for off-earth mining, and open new avenues for expanding the space economy. Also, a new level of commercial and international partnership opportunities are opened which create the sense of global cooperation and common human endeavors. In the long-term vision of deep space exploration--particularly in mankind's journey to Mars--a permanent facility on the Moon can be the launching pad and gateway for deep space communication and commercial activities. A key challenge of Artemis' success will be a reliable and sustainable End-to-End (E2E) communication network~\cite{LunaNET_2020}. Various technologies have been proposed for E2E communication links between ground stations, lunar orbiters (Gateways), and lunar rovers~\cite{LunaNET_2020} focusing on architectural design, lunar constellations for far side explorations using disruption-tolerant protocols~\cite{LunaNET_2020}, and commercial and international partnership opportunities.
\par Communication in traditional lunar networks has overwhelmingly been based on Radio Frequency (RF) links, which provide reliable communication but are inherently energy inefficient--a serious concern for power-limited Small Satellites (SmallSats). Contrary to this, optical communication can provide better energy efficiency and higher data rates, but susceptibility to unfavorable weather conditions and Pointing and Tracking (PAT) errors limit its widespread terrestrial use. However, several studies have shown that Free Space Optical (FSO) communication provides a viable option for future lunar communication (LunarComm) and space communications studies~\cite{2017_LCRD}. Therefore, a practical approach is to design a future communication system that is flexible, integrated, and adaptive, which can reap the benefits of both paradigms. Hybrid RF/FSO systems have been widely studied in the paradigm of terrestrial networks and Low Earth Orbit (LEO) satellites~\cite{2021_Hybrid_HAPS}. NASA's integrated Radio Optical Communication (iROC) focused on developing a single terminal based hybrid RF and optical communication for space communication~\cite{iROC_2017}. \textit{However, to the best of the authors' knowledge there is currently no demonstrated SmallSat and space networking strategy for lunar networks which utilizes both RF and FSO links.}
\begin{figure*}[t!]
	\centering
	\includegraphics[width=0.8\textwidth, height=6cm]{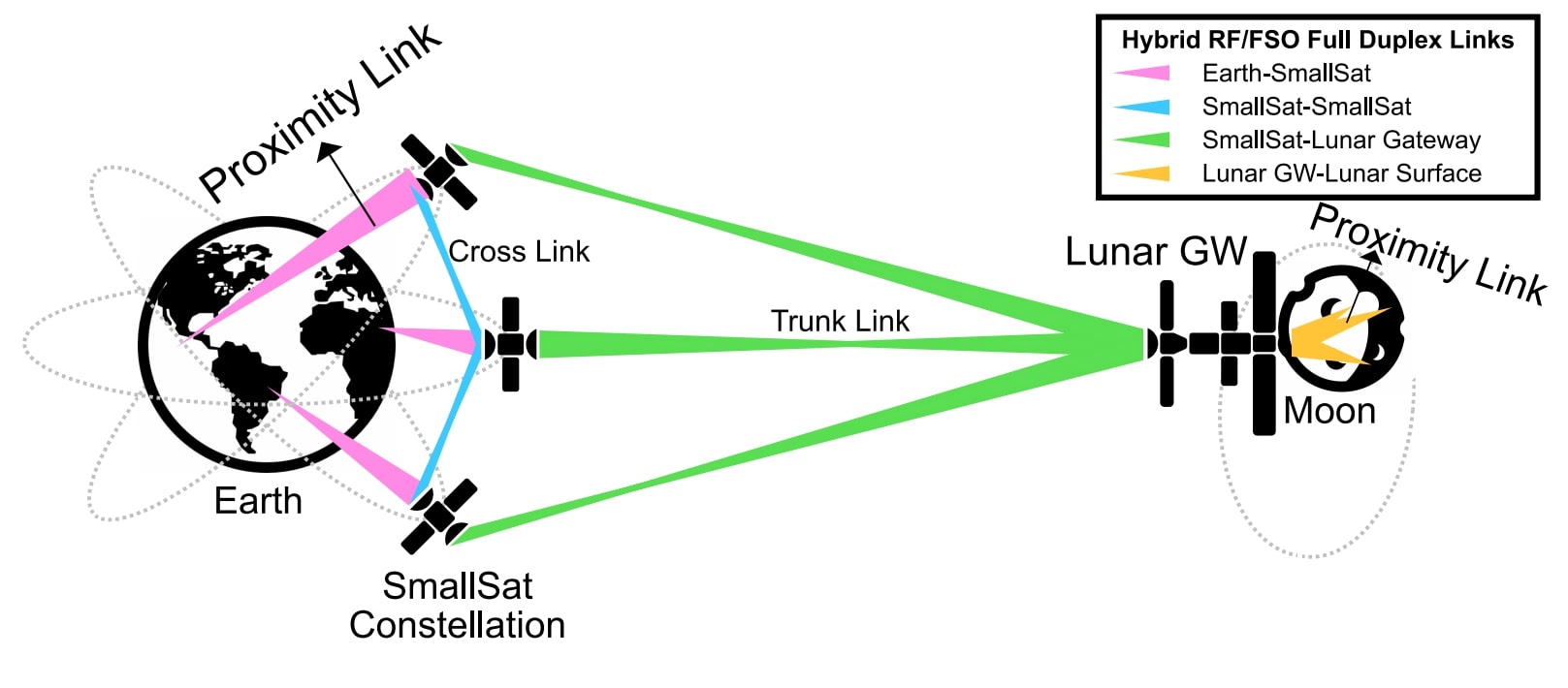}
	\caption{Schematic diagram of a hybrid lunar communication network.}
	\label{fig:architecture}
\end{figure*}
\par Motivated by this gap, we argue for a hybrid RF/FSO-based lunar network design which utilizes a SmallSat based LEO constellation where each satellite is equipped with both RF and FSO transceivers to communicate with earth stations and the Lunar Gateway (GW), which also has similar transceivers onboard. The contribution is summarized in the following.
\begin{itemize}
\item {We first discuss the two-hop general architecture of LunarComm, which includes the constellations of SmallSats in LEO orbit, Lunar GW, and facilities on Earth and Lunar Surfaces. Our preliminary hybrid solution includes \textit{hard switching} between RF and FSO links where the RF links have primary role of supporting the low data rates, while switching to FSO link is performed for higher data rates. Also, RF links are used to carry the control data for the initial FSO signal acquisition and tracking.}
\item {Then, the important features of RF and FSO based communication systems are discussed with focus on their strength and weaknesses. The comparative limitations and benefits of each scheme suggest that a hybrid combination will provide the best of both approaches, and various aspects of the hybrid solutions are discussed.}
\item {Next, the proposed Hybrid RF/FSO system is discussed subsequently followed by a discussion on the relevance of proposed work with NASA's vision.}
\item {We design a case study using the communication module of the Analytical Graphics, Inc (AGI) Systems Tool Kit (STK) to demonstrate the benefits of a hybrid solution in terms of propagation loss, Bit Error Rate (BER) and Signal-to-Noise Ratio (SNR). We deploy various constellations of SmallSats to analyze the coverage in terms of access duration at earth facility.} 
\end{itemize}
\section{General Architecture for the LunarComm}
The general architecture of the proposed LunarComm Network is shown Fig.~{\ref{fig:architecture}}. It leverages two intermediate relays, i.e., constellations of satellites on the Earth side and a GW on the lunar side to assist the transmission from Earth to Lunar facilities. The links from the SmallSats to Earth and from GW to moon are termed ``proximity links'' and between SmallSats and GW as ``trunk links''. The satellites in the constellations can communicate with each other over the cross links. Both the trunk and proximity links can be switched between RF and FSO communication. Hence the respective devices are equipped with both type of transceivers. For RF and FSO channel models we resort to the basic models available in AGI STK, i.e., ITU-R P.168 Section 2.5 for RF and ITU-R P1814 for FSO communication. Further, the communication system is designed in the AGI STK Communication module, which follows the theoretical model to calculate the access and communication results \footnote{{The values of various parameters are obtained from the state-of-the-art works in Lunar Laser Communication Demonstration, Laser Communications Relay Demonstration, and Inter-agency Operations Advisory Group (IOAG) documentation.}}. The orbital height and constellations of satellites depend on various factors, such as the PAT error, delay and coverage requirements in trunk and proximity links, and cost and energy consumption constraints of the satellites. Satellites in GEO orbit can offer reliable connection and broad coverage because of their stationary positions to the earth rotation and relatively less distance from the Lunar surface or Gateway. This can also bring some relaxation in PAT requirements. However, it will incur increased delay and energy consumption on earth side proximity link. On the other hand, the satellite in LEO orbit, because of the proximity to the Earth, can be a better choice, especially for energy-constrained SmallSats. The NASA Artemis program includes the GW as a vital component for sustainable lunar explorations. Similar to the International Space Station, it will be a solar-powered communication hub, autonomously operating, science laboratory and short term habitation module for astronauts. The Lunar GW can be utilized as an important intermediate relay to communicate to the lunar surface, and because of its deployment in near rectilinear halo orbit, its line of sight to the Earth would be minimally obstructed.

\section{RF and FSO Communication Overview}
\subsection{Radio Frequency Communication}
The greatest strengths of RF communication are its ability to penetrate through most atmospheric conditions and immunity to precise alignment requirements. Modern RF space communications such as the onboard NASA's latest Tracking and Data Relay Satellite (TDRS) use transceivers operating primarily in the S, X, and K bands \cite{TDRS_X_Ka} having frequencies from {$2$} to {$27$} GHz and relying on line-of-sight propagation between transmitters and receivers. The {K\textsubscript{a}} band ({$26.5$} to {$40$} GHz) has gained much interest in recent years due to its higher achievable data rates and greater spectrum availability. The TDRS has demonstrated data rates up to {$800$} Mbps while using {K\textsubscript{a}}-band links {\cite{TDRS_X_Ka}}. Off-the-shelf hardware designed for the {K\textsubscript{a}} band is also becoming more readily available as applications proliferate.

Regarding the modulation and coding, most of the current and planned space missions are designed around simpler schemes such as Phase Shift Keying (PSK)~{\cite{Update_CCSDS_optical}}. {Space applications often benefit from choosing coding schemes and rates separately for uplink and downlink, and an adaptive coding rate can be useful for changing atmospheric conditions. Coding rates of $1/2$, $2/3$, and $4/5$ alongside convolution or Reed-Solomon coding schemes have performed well in space applications}~{\cite{2020_FSOC_CubeSats_LLO,Update_CCSDS_optical}}. {Recent advancements in Software Defined Radio (SDR) technology present an attractive option for building space communication systems. SDRs not only allow for rapid development with flexible modulation and carrier selections but also permit many post-deployment alterations in link parameters. In space applications, this capability allows engineers to adaptively update the link specifications for the SDR-based relay satellites in-flight}~{\cite{2015_SDR_Small_Sats}}. SDRs are becoming smaller and lighter, and off-the-shelf units can be purchased for small satellite applications. Finally, RF technology provides backwards compatibility. RF's dominance in past decades means that many existing and legacy space systems can only be contacted via RF. Therefore, RF technology must be included on any system that must communicate with previous generations of technologies.

{The primary weakness of RF technology and the motivation for moving to FSO is its comparatively low data rates. FSO has demonstrated potential for orders-of-magnitude greater data rates. The next generation of space communications will require high-throughput links to carry data for various public and commercial missions, and RF alone is not currently up to that task. RF links also require a significant amount of room onboard spacecraft for each antenna, which can also be relatively heavy. Antenna technology for SmallSats has advanced, but with in-flight deployment methods rather than weight reduction as the primary focus}~{\cite{TDRS_X_Ka}}. {In the {K\textsubscript{a}} band specifically, there also exists a greater concern of atmospheric effects. The {K\textsubscript{a}} band's higher achievable data rates come with a compromise known as rain fading--the attenuation of an RF signal due to absorption by water droplets in the atmosphere. This effect is most prevalent in frequencies above $11$ GHz, which causes the limited consideration in existing S- and X-band links}~{\cite{TDRS_X_Ka}}.

\subsection{Free Space Optical Communication}
{Optical communication addresses some of the important limitations of RF systems because optical antennas can produce a higher signal gain in a much smaller footprint (often around $10$ -- $20$ cm) and use lighter-weight components (plastics, glass). FSO-based communication offers much higher data rates, which are critical to increase the capabilities of telemetry, science, voice, video, and alerts}~\cite{Update_CCSDS_optical}. {Optical communication systems provide greater immunity to interference from other sources than RF systems, as they would not interfere with GEO communication (an ITU requirement) nor any other RF systems such as the high data rate RF links envisioned by the IOAG.}

{Nevertheless, FSO systems also benefit from existing technology and research. Many of the components needed for advanced, high-speed optical communication exist as commercial off-the-shelf components developed for fiber optic systems. A wealth of components and technology exist to support communication at $1550$-nm wavelengths recommended for lunar and lunar-to-earth forward and return links in the Consultative Committee for Space Data Systems (CCSDS)}~{\cite{Update_CCSDS_optical}}. {Many components have already undergone reductions in size, weight, and power consumption critical to the operation of satellites, Lunar GWs, and other deep-space assets through the development of integrated photonics and opto-electronic integrated circuits.}

{FSO systems face several challenges beyond the atmospheric and mechanical effects. Link budget analyses estimate that an optical transmitter power of $10-100$ W is needed to effectively detect high data rate signals at earth-to-moon distances--a difficult goal for power-limited SmallSats. A trade-off exists between system efficiency and data rate, with efficiency seen as more important on deep space assets since they are so power limited,}~{\cite{Update_CCSDS_optical}} {and high data rate seen as a higher priority for LEO except in cases such as SmallSats or CubeSats that are far more power-limited. It is expected that the underlying technologies and techniques for modulation, coding, and synchronization will be significantly different between the two signal cases}~{\cite{Update_CCSDS_optical}}. {Some other concerns in FSO include the effects of the blast ejecta (lunar dust) on optical surfaces, acquisition and tracking, and interoperability with existing or planned communication assets.}

\subsection{Hybrid RF/FSO Constellation}
{From the discussion in the previous section, it is emphasized that the RF and FSO techniques have complimentary features. Both RF and FSO communication technologies are well studied in the literature. However, FSO technology is significantly less mature, and is currently relegated to demonstration projects such as NASA's CubeSat Laser Infrared Crosslink (CLICK) or the Laser Communication Relay Demonstration. A hybrid system can balance the high throughput of FSO with the reliability/wide coverage of RF links to maximize performance in challenging design environments such as the E2E lunar communication system. In fact, similar methodology has been used to improve other systems such as $5G$ backhaul links}~\cite{2017_PIMRC_Hybrid_5G}. {There is also active work in hybrid RF/FSO communication strategies for ground to satellite communications}~\cite{2021_Hybrid_HAPS}. {However, the authors are aware of no existing effort to combine these technologies to achieve Earth-to-Lunar communications using Lunar Gateway and SmallSat constellations. Such a system has several important advantages compared to the standard methodology of direct communication with ground stations on earth. Earth's atmosphere introduces undesirable effects, particularly in FSO communications. These include scintillation and diffraction, among others. Such effects are magnified by distance, even after an optical signal has left the atmosphere. Intercepting signals from ground stations in LEO minimizes these effects, and the signal can be immediately re-transmitted free from atmospheric distortions. The availability of RF links in this constellation may serve as a backup in adverse weather conditions and may also be used to handle telemetry and command data for the constellation satellites while leaving the optical links free for communication data.}

{While the benefits of hybrid RF/FSO systems are clear, the realization of a truly hybrid system is a much more involved process requiring further research. A simple hybrid network could be built by designing a system with both RF and FSO transceivers mounted on the same device with each system having separate transmitting and receiving chains. Such a system would require a hard switching and feedback mechanism between transmitter and receiver to jointly select the RF or FSO link for communication. By contrast, soft-switching based hybridization utilizes: (1) channel coding (mainly Low Density Parity Check and Raptor codes) with a portion of the codeword split to RF and FSO links, (2) channel conditions-based puncturing for rate adjustments, and (3) complex soft decoding at FSO supported data rates}~\cite{2009_Hard_Soft}. {Although soft switching utilizes the common modulation and encoding modules, its RF antenna and FSO laser are separately mounted. A more sophisticated approach is to design a hybrid antenna with adaptable apertures to operate on both the RF and FSO links as discussed in}~\cite{2016_Dual_Antenna}.

\subsection{Relevance to NASA Future Vision}
{The proposed hybrid RF/FSO architecture would be important for NASA's Artemis mission as well as for future networks around other planetary bodies such as Mars. 
For example, NASA is currently working towards developing the ``LunaNet" architecture that aims to ``empower Artemis with communications and navigation interoperability"}\cite{LunaNET_2020}. {Considering the high-rate and reliability needs of such a communication network, the proposed LunarComm architecture shows the strong alignment with LunaNet.} 

{In addition, the future LunarComm network must have both high data rate and high reliability despite an unprecedented series of challenges. The proposed constellation of SmallSats greatly increases the redundancy, reliability and accessibility of the network compared to direct Earth-Lunar communication architectures. SpaceX's Starlink network has demonstrated the feasibility of world-wide network coverage using LEO SmallSats. The architecture proposed here will apply this same accessibility to Lunar communications. Any university or institution may directly access the network from any location. This accessibility also creates improved reliability and scalability. Critical missions with large budgets may build or contract multiple optical ground stations locations across the globe to ensure optical data rates at all times. Less critical or budget-limited missions may rely on only a single ground station, while still benefiting from the RF backup to maintain basic command and telemetry readings. 
 
These proposed advantages can be enabled both by advances in hardware and in network strategies. The SmallSat nodes must be capable of communication through atmosphere, inter-satellite crosslink, and LEO-to-Lunar communication. Each of these requirements creates different challenges. Satellite crosslink and LEO-to-Lunar links do not suffer the challenges of atmospheric effects but may require much greater pointing accuracy given the larger distances. Meeting each of these challenges simultaneously will require innovative designs given the limited SWaP requirements of SmallSat platforms. Intelligent switching strategies must also be employed to optimize network throughput in varying conditions. SmallSat nodes may become inaccessible for a variety of reasons, ranging from poor atmospheric conditions to low power. The network must manage throughput in a constantly changing environment.}

\begin{table}[b]
\setlength{\tabcolsep}{1pt}
\centering
\caption{{Orbital parameters of the LEO Sat and Lunar GW.}}
\label{tab:orb_params}
\begin{tabular}{|c|c|c|c|c|c|c|c|}
	\hline
	\textbf{Parameters} &
	\textbf{\begin{tabular}[c]{@{}c@{}}Coord\\inate\\ Type\end{tabular}} &
	\textbf{\begin{tabular}[c]{@{}c@{}}Semi\\ Major \\ Axis \end{tabular}} &
	\textbf{\begin{tabular}[c]{@{}c@{}}Eccen\\ tricity\end{tabular}} &
	\multicolumn{1}{l|}{\textbf{\begin{tabular}[c]{@{}l@{}}Incli\\nation\end{tabular}}} &
	\textbf{\begin{tabular}[c]{@{}c@{}}Arg.\\ of \\ Perg.\end{tabular}} &
	\textbf{\begin{tabular}[c]{@{}c@{}}Long. \\ of \\ Ascend. \\Node\end{tabular}} &
	\textbf{\begin{tabular}[c]{@{}c@{}}True \\ Ano\\maly\end{tabular}} \\ \hline
	\textbf{\begin{tabular}[c]{@{}c@{}}Initial Small\\ Sat Orbit\end{tabular}} &
	Classical &
	\begin{tabular}[c]{@{}c@{}}7400\\ (1000) km\end{tabular} &
	0 &
	120 &
	0 &
	0 &
	90 \\ \hline
	\textbf{\begin{tabular}[c]{@{}c@{}}Lunar \\ Gateway\end{tabular}} &
	Classical &
	6142.4 km &
	0.6 &
	67.7 &
	270 &
	270 &
	90 \\ \hline
\end{tabular}
\end{table}
\begin{table}[b]
\setlength{\tabcolsep}{1pt}
\centering
\caption{{Access Statistics for different Constellations.}}
\label{tab:constellationslation_access}
	\begin{tabular}{|c|c|c|c|c|c|}
		\hline
		\textbf{Constellation} & \textbf{Count} & \textbf{Min} & \textbf{Mean} & \textbf{Max} & \textbf{Sum} \\ \hline
		\textbf{1 by 1} & 9  & 150.63   & 800.57   & 1026.86  & 7205.13  \\ \hline
		\textbf{2 by 2} & 29 & 150.63   & 940.42   & 1185.90  & 27272.04 \\ \hline
		\textbf{3 by 3} & 43 & 499.10   & 1361.58  & 1763.17  & 58547.84 \\ \hline
		\textbf{4 by 4} & 1  & 86400.00 & 86400.00 & 86400.00 & 86400.00 \\ \hline
	\end{tabular}
\end{table}

\section{Hybrid RF/FSO End-to-End Lunar Network Case Study}
\begin{figure}[t]
	\centering
	\begin{subfigure}{0.44\textwidth}
		\centering
		\includegraphics[width=8cm, height=3cm]{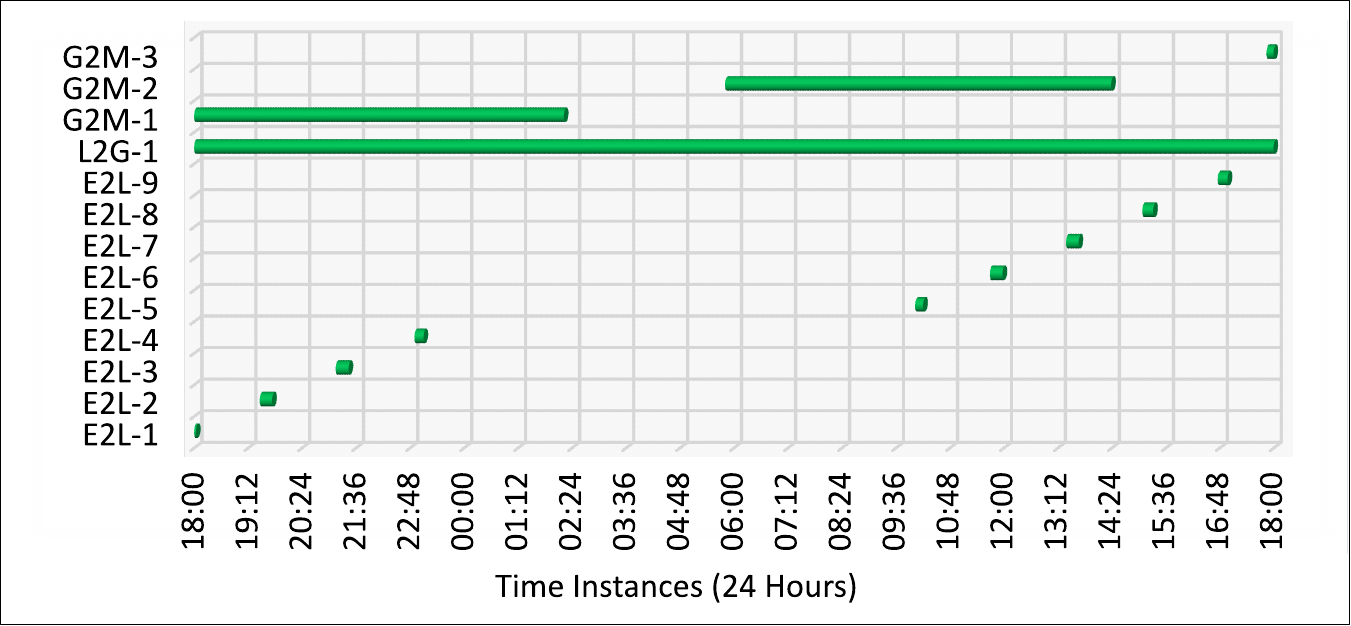}
        \caption{{E2L access durations for 1 by 1 constellation.}}
		\label{fig:chain_access_time}
	\end{subfigure}
~
	\begin{subfigure}{0.22\textwidth}
		\centering
		\includegraphics[width=4cm, height=3cm]{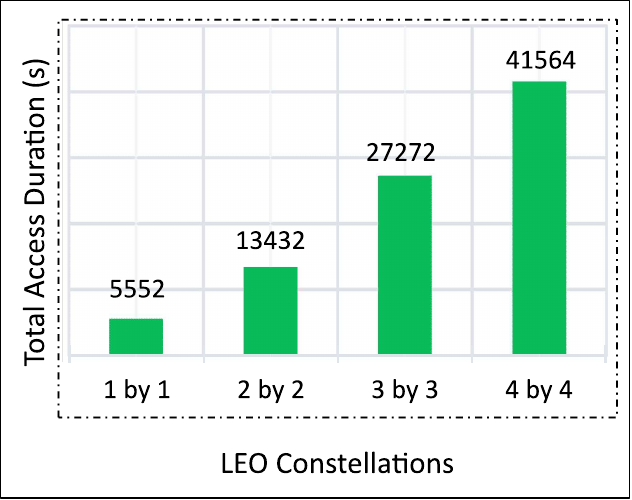}
        \caption{{End-to-end access duration for different constellations.}}
		\label{fig:access_time}
	\end{subfigure}
	\begin{subfigure}{0.22\textwidth}
		\centering
		\includegraphics[width=4cm, height=3cm]{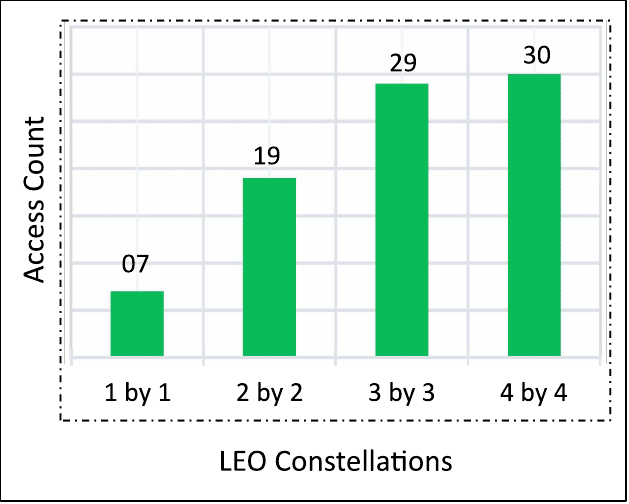}
		\caption{{End-to-end access count for different constellations.}}
		\label{fig:access_count}
	\end{subfigure}
\caption{Chain access results for LEO Satellite Constellations.}
\label{fig:chain_accesses}
\vspace{-5mm}
\end{figure}

\vspace{-2mm}
In this section, we design and evaluate the proposed hybrid RF/FSO based network for E2E communication between the earth and moon facilities. We consider a single earth station as the communication source and a single lunar facility as the target, and utilize the AGI STK Communication module for implementing the E2E link with both RF and FSO components. Initially a single LEO satellite with orbital parameters given in~Table~\ref{tab:orb_params} is deployed, however, its significantly larger invisibility duration from the earth station, as given in Fig.~\ref{fig:chain_access_time}, suggests that a constellation of satellites should be used instead. Hence, we take the initial LEO orbit as the basic model of the Walker constellation tool. This keeps the orbits and number of satellites in each orbit at maximum separation, and creates different constellations, i.e., \textit{2-by-2}, \textit{3-by-3} and \textit{4-by-4}, which collectively act as the relay station. {Aiming to achieve the maximum possible visibility by utilizing the minimum possible constellations, we start observing the access results for \textit{1-by-1} and note the improvements until 100\% visibility is realized.} On the lunar side, we deploy GW which also acts as the second relay node for the E2E communication chain from Earth station to any Lunar facility~\cite{LunaNET_2020}. Both relays are equipped with separate sensors for aligning towards each other and with their respective communication source or destination. Each sensor is equipped with both RF and FSO transceivers. The orbital parameters of the GW are given in~Table~\ref{tab:orb_params} and values of different parameters of RF and FSO chains follow from the standards\cite{Update_CCSDS_optical}, as given in Table\textbf{~\ref{tab:orb_params}}. The RF and FSO components are mounted on the physical assets and these are assumed to be inter-operable and compatible with each other. We consider that the RF link is always present in the whole communication system to provide reliable low data rate communication, however, hard switching from RF to FSO links on each ``hop" of the network is possible to support the high data rate transmission. The access/visibility results of the initial single satellite based deployment and subsequent extension to constellations are discussed in the following subsection.
\subsection{Individual and Chain Access Results}
Visibility between the deployed assets is an important parameter in supporting sustainable communication performance. However this visibility varies because of the relative motion of the devices. In Fig.~\ref{fig:chain_access_time}, we present the access duration between different pairs of individual objects for a period of $24$ hours. The figure gives a detailed depiction of each visibility instance i.e., E2L (accesses 1 through 9): Earth and a SmallSat in LEO orbit (for the 1-by-1 constellation case), L2G: LEO to GW, and G2M (accesses 1 through 3): GW to the lunar facility. The total $9$ instances of the access (disruptions in visibility) between Earth and SmallSat, accumulate a total time of $7205.132$ seconds or $8.34\%$ of $24$ hours. LEO and GW are placed in such positions that they are visible to each other for the whole duration of the $24$ hours. The GW to lunar visibility is affected by three disruptions, with total access time of $61068.353$ or $70.68\%$. 

It can be concluded that the access between Earth station and LEO is the key factor to maintain the E2E visibility from E2L stations. Hence, in Table~\ref{tab:constellationslation_access}, we focus on this first hop link (E2L) and observe the visibility statistics by increasing the number of orbits and satellites per orbit. Thus the total visibility duration between Earth facility and any satellite in \textit{2-by-2} constellations is increased to $27272.037$ seconds or $31.56\%$ of one day's time. Similarly, for \textit{3-by-3} constellations visibility is improved to $67.7\%$ and we get the $100\%$ visibility for \textit{4-by-4} LEO constellation. This means that at any given time at least one of $16$ deployed satellites is in view of the Earth facility. {From these results, it is generally observed that the access duration steadily increases with the scale of LEO satellite constellations. Hence, for increased reliability and continuous coverage the use of more satellites and denser constellations is an intuitive option. However, apart from the increased cost, there will be increased disruptions and switching/hopping with satellites, and sophisticated multiple access techniques would be required to ensure maximum uninterrupted coverage. Also, it will enhance the inter-satellite interference and require better resource management techniques for conflict avoidance.} 

Table \ref{tab:constellationslation_access} demonstrated the improvement in access within the first hop link (E2L) only as constellation satellites become more numerous. However, the \emph{total chain} access is equally important. {Hence, we present the total duration of E2E chain accesses for different combinations of satellite constellations in Fig.}~\ref{fig:access_time}. This result show a general trend that the total chain access increases from $6.42\%$ to $48.10\%$ when the constellations are increased from \textit{1-by-1} to \textit{4-by-4}. Although, the use of \textit{4-by-4} constellations has provided full time connectivity from the constellation to the Earth facility, the chain access still could not achieve full coverage because of the limitations from the GW to Lunar link. One possible approach to improve this connectivity would be to introduce a constellation of Gateways, having at least one additional Gateway in the other halo orbit, but we skip this analysis for simplicity. The trend of the access count shown in Fig.~\ref{fig:access_count} is also similar to the duration of access for different constellations. It should be noted that the higher access counts are not always beneficial, since they represent the disruption and reestablishment of connection.
\begin{table*}[tbh]
\setlength{\tabcolsep}{2pt}
\centering
\caption{LunarComm Case Study: RF and FSO input parameters and performance comparison with transmit power of 1 dBW.}
\label{tab:params_values}
\begin{tabu}{|[1pt]c|[1pt]c|[1pt]ccccccccc|[1pt]cccccc|[1pt]c|[1pt]c|[1pt]}
\tabucline[1pt]{1-19}
\multirow{5}{*}{\textbf{\begin{tabular}[c]{@{}c@{}}Link \\ Direction\end{tabular}}} &
  \multirow{5}{*}{\textbf{\begin{tabular}[c]{@{}c@{}}Device \\ Name\end{tabular}}} &
  \multicolumn{9}{c|[1pt]}{\textbf{RF Link   Information}} &
  \multicolumn{6}{c|[1pt]}{\textbf{FSO Link   Information}} &
  \multirow{5}{*}{\textbf{\begin{tabular}[c]{@{}c@{}}Distance \\ (Km)\end{tabular}}} &
  \multirow{5}{*}{\textbf{\begin{tabular}[c]{@{}c@{}}Prop. \\ Delay \\ (s)\end{tabular}}} \\ \cline{3-17}
 &
   &
  \multicolumn{6}{c|}{\textbf{RF Input Parameters}} &
  \multicolumn{3}{c|[1pt]}{\textbf{RF Link BER}} &
  \multicolumn{3}{c|}{\textbf{FSO Input Parameters}} &
  \multicolumn{3}{c|[1pt]}{\textbf{FSO Link BER}} &
   &
   \\ \cline{3-17}
 &
   &
  \multicolumn{1}{c|}{\textbf{\begin{tabular}[c]{@{}c@{}}Oper. \\ Freq \\ (GHz)\end{tabular}}} &
  \multicolumn{1}{c|}{\textbf{\begin{tabular}[c]{@{}c@{}}Anten. \\ Diam\\  (m)\end{tabular}}} &
  \multicolumn{1}{c|}{\textbf{\begin{tabular}[c]{@{}c@{}}Beam\\ width \\ (deg)\end{tabular}}} &
  \multicolumn{1}{c|}{\textbf{\begin{tabular}[c]{@{}c@{}}Gain \\ (dB)\end{tabular}}} &
  \multicolumn{1}{c|}{\textbf{\begin{tabular}[c]{@{}c@{}}Data\\  Rate \\ (Mbps)\end{tabular}}} &
  \multicolumn{1}{c|}{\textbf{Modul}} &
  \multicolumn{1}{c|}{\textbf{min}} &
  \multicolumn{1}{c|}{\textbf{mean}} &
  \textbf{max} &
  \multicolumn{1}{c|}{\textbf{\begin{tabular}[c]{@{}c@{}}Ae \\ (m$^2$)\end{tabular}}} &
  \multicolumn{1}{c|}{\textbf{\begin{tabular}[c]{@{}c@{}}Gain \\ (dB)\end{tabular}}} &
  \multicolumn{1}{c|}{\textbf{\begin{tabular}[c]{@{}c@{}}Data \\ Rate \\ (Mbps)\end{tabular}}} &
  \multicolumn{1}{c|}{\textbf{min}} &
  \multicolumn{1}{c|}{\textbf{mean}} &
  \textbf{max} &
   &
   \\ \tabucline[1pt]{1-19}
\multirow{6}{*}{\textbf{\begin{tabular}[c]{@{}c@{}}Up\\ Link \\ \\ Earth \\ to \\ Moon \\ Link\end{tabular}}} &
  \textbf{\begin{tabular}[c]{@{}c@{}}Earth \\ UL Tx\end{tabular}} &
  \multicolumn{1}{c|}{10} &
  \multicolumn{1}{c|}{1} &
  \multicolumn{1}{c|}{2.32} &
  \multicolumn{1}{c|}{37.81} &
  \multicolumn{1}{c|}{15} &
  \multicolumn{1}{c|}{QPSK} &
  \multicolumn{1}{c|}{\multirow{2}{*}{\begin{tabular}[c]{@{}c@{}}2.38\\ E-07\end{tabular}}} &
  \multicolumn{1}{c|}{\multirow{2}{*}{\begin{tabular}[c]{@{}c@{}}1.74\\ E-05\end{tabular}}} &
  \multirow{2}{*}{\begin{tabular}[c]{@{}c@{}}1.09\\ E-04\end{tabular}} &
  \multicolumn{1}{c|}{0.01} &
  \multicolumn{1}{c|}{100} &
  \multicolumn{1}{c|}{1000} &
  \multicolumn{1}{c|}{\multirow{2}{*}{\begin{tabular}[c]{@{}c@{}}2.04\\ E-15\end{tabular}}} &
  \multicolumn{1}{c|}{\multirow{2}{*}{\begin{tabular}[c]{@{}c@{}}4.84\\ E-07\end{tabular}}} &
  \multirow{2}{*}{\begin{tabular}[c]{@{}c@{}}7.46\\ E-08\end{tabular}} &
  \multirow{2}{*}{7725.42} &
  \multirow{2}{*}{0.019} \\ \cline{2-8} \cline{12-14}
 &
  \textbf{\begin{tabular}[c]{@{}c@{}}LEO \\ UL Rx\end{tabular}} &
  \multicolumn{1}{c|}{10} &
  \multicolumn{1}{c|}{0.25} &
  \multicolumn{1}{c|}{9.26} &
  \multicolumn{1}{c|}{25.77} &
  \multicolumn{1}{c|}{-} &
  \multicolumn{1}{c|}{-} &
  \multicolumn{1}{c|}{} &
  \multicolumn{1}{c|}{} &
   &
  \multicolumn{1}{c|}{0.01} &
  \multicolumn{1}{c|}{105} &
  \multicolumn{1}{c|}{-} &
  \multicolumn{1}{c|}{} &
  \multicolumn{1}{c|}{} &
   &
   &
   \\ \tabucline[1pt]{2-19} 
 &
  \textbf{\begin{tabular}[c]{@{}c@{}}LEO \\ UL Tx\end{tabular}} &
  \multicolumn{1}{c|}{34} &
  \multicolumn{1}{c|}{0.75} &
  \multicolumn{1}{c|}{0.91} &
  \multicolumn{1}{c|}{45.94} &
  \multicolumn{1}{c|}{1} &
  \multicolumn{1}{c|}{\begin{tabular}[c]{@{}c@{}}BPSK\\ -BCH-\\ 127-64\end{tabular}} &
  \multicolumn{1}{c|}{\multirow{2}{*}{\begin{tabular}[c]{@{}c@{}}1.00\\ E-10\end{tabular}}} &
  \multicolumn{1}{c|}{\multirow{2}{*}{\begin{tabular}[c]{@{}c@{}}6.51\\ E-08\end{tabular}}} &
  \multirow{2}{*}{\begin{tabular}[c]{@{}c@{}}1.22\\ E-07\end{tabular}} &
  \multicolumn{1}{c|}{0.05} &
  \multicolumn{1}{c|}{112} &
  \multicolumn{1}{c|}{300} &
  \multicolumn{1}{c|}{\multirow{2}{*}{\begin{tabular}[c]{@{}c@{}}9.57\\ E-12\end{tabular}}} &
  \multicolumn{1}{c|}{\multirow{2}{*}{\begin{tabular}[c]{@{}c@{}}1.28\\ E-11\end{tabular}}} &
  \multirow{2}{*}{\begin{tabular}[c]{@{}c@{}}1.57\\ E-11\end{tabular}} &
  \multirow{2}{*}{364465} &
  \multirow{2}{*}{1.216} \\ \cline{2-8} \cline{12-14}
 &
  \textbf{\begin{tabular}[c]{@{}c@{}}GW \\ UL Rx\end{tabular}} &
  \multicolumn{1}{c|}{34} &
  \multicolumn{1}{c|}{1.25} &
  \multicolumn{1}{c|}{1.36} &
  \multicolumn{1}{c|}{50.38} &
  \multicolumn{1}{c|}{-} &
  \multicolumn{1}{c|}{-} &
  \multicolumn{1}{c|}{} &
  \multicolumn{1}{c|}{} &
   &
  \multicolumn{1}{c|}{0.05} &
  \multicolumn{1}{c|}{112} &
  \multicolumn{1}{c|}{-} &
  \multicolumn{1}{c|}{} &
  \multicolumn{1}{c|}{} &
   &
   &
   \\ \tabucline[1pt]{2-19} 
 &
  \textbf{\begin{tabular}[c]{@{}c@{}}GW\\  UL Tx\end{tabular}} &
  \multicolumn{1}{c|}{10} &
  \multicolumn{1}{c|}{0.5} &
  \multicolumn{1}{c|}{4.63} &
  \multicolumn{1}{c|}{31.79} &
  \multicolumn{1}{c|}{15} &
  \multicolumn{1}{c|}{QPSK} &
  \multicolumn{1}{c|}{\multirow{2}{*}{\begin{tabular}[c]{@{}c@{}}1.23\\ E-09\end{tabular}}} &
  \multicolumn{1}{c|}{\multirow{2}{*}{\begin{tabular}[c]{@{}c@{}}1.37\\ E-08\end{tabular}}} &
  \multirow{2}{*}{\begin{tabular}[c]{@{}c@{}}4.55\\ E-08\end{tabular}} &
  \multicolumn{1}{c|}{0.01} &
  \multicolumn{1}{c|}{100} &
  \multicolumn{1}{c|}{1000} &
  \multicolumn{1}{c|}{\multirow{2}{*}{\begin{tabular}[c]{@{}c@{}}1.62\\ E-14\end{tabular}}} &
  \multicolumn{1}{c|}{\multirow{2}{*}{\begin{tabular}[c]{@{}c@{}}8.63\\ E-13\end{tabular}}} &
  \multirow{2}{*}{\begin{tabular}[c]{@{}c@{}}8.63\\ E-13\end{tabular}} &
  \multirow{2}{*}{8729.26} &
  \multirow{2}{*}{0.025} \\ \cline{2-8} \cline{12-14}
 &
  \textbf{\begin{tabular}[c]{@{}c@{}}Lunar \\ UL Rx\end{tabular}} &
  \multicolumn{1}{c|}{10} &
  \multicolumn{1}{c|}{0.5} &
  \multicolumn{1}{c|}{4.63} &
  \multicolumn{1}{c|}{31.79} &
  \multicolumn{1}{c|}{-} &
  \multicolumn{1}{c|}{-} &
  \multicolumn{1}{c|}{} &
  \multicolumn{1}{c|}{} &
   &
  \multicolumn{1}{c|}{0.01} &
  \multicolumn{1}{c|}{100} &
  \multicolumn{1}{c|}{-} &
  \multicolumn{1}{c|}{} &
  \multicolumn{1}{c|}{} &
   &
   &
   \\ \tabucline[1pt]{1-19}\hline
\end{tabu}
\end{table*}
\begin{figure}[b]
	\centering
	\begin{subfigure}[b]{0.48\linewidth}
		\centering
		\includegraphics[width=4cm, height=3cm]{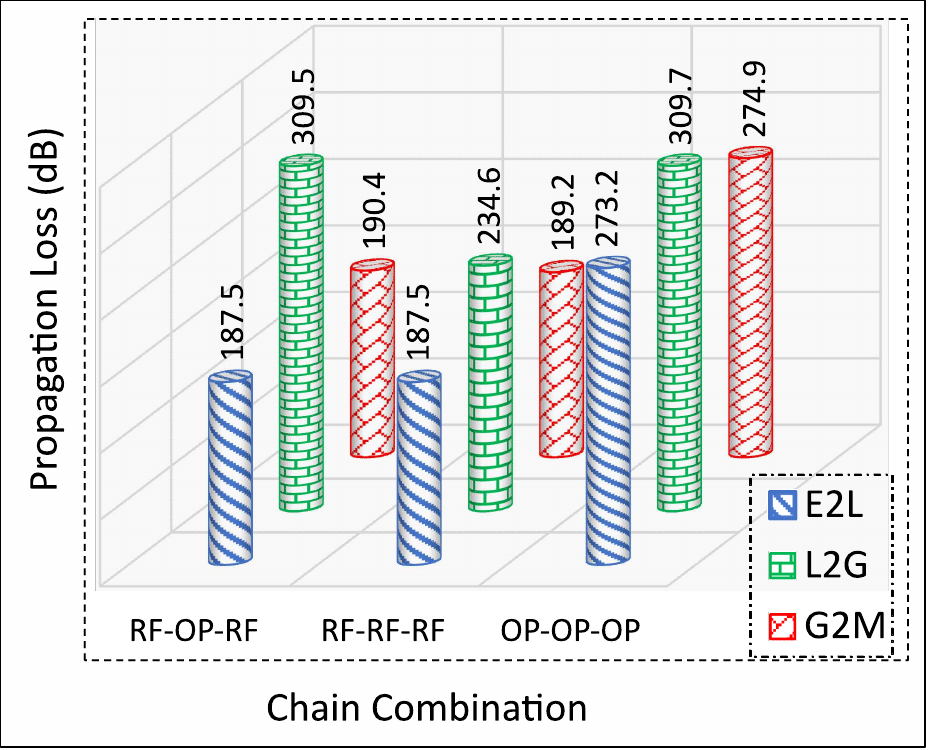}
	\end{subfigure}	
	\begin{subfigure}[b]{0.48\linewidth}
		\centering
		\includegraphics[width=4cm, height=3cm]{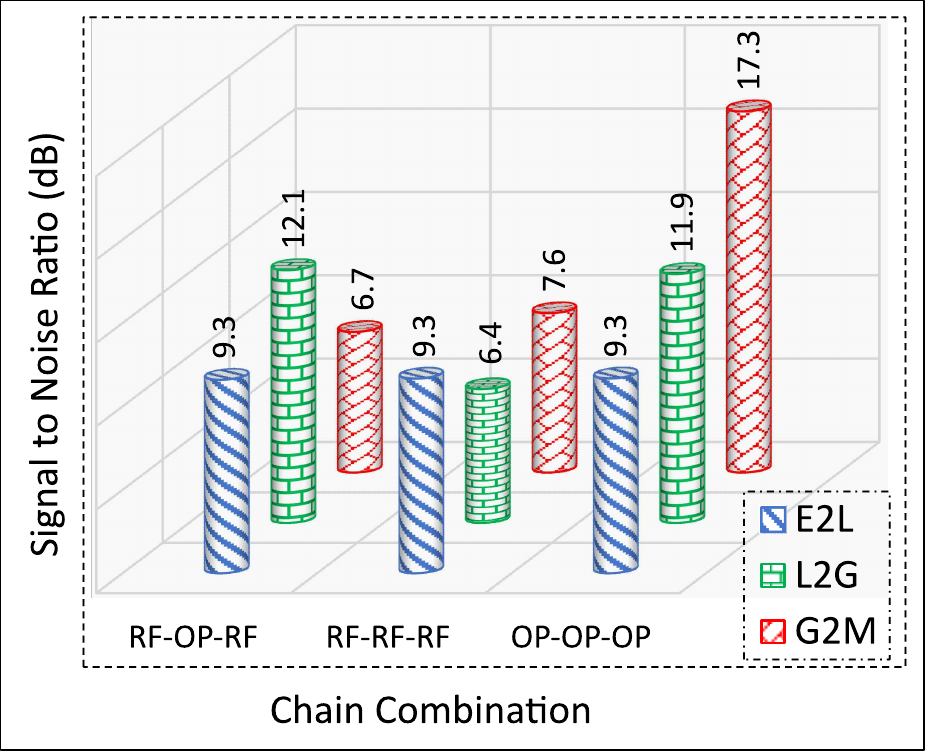}
	\end{subfigure}
	\caption{{Hybrid RF/FSO based Lunar Communication Performance Comparison for propagation loss (left) and SNR (right).}}
	\label{fig:prop_loss_snr}
\end{figure}
\subsection{Comparing the Hybrid Solution to Individual Counterparts}
For the link budget analysis we design two different chains in AGI STK Communication for the RF and FSO links as detailed in Table~\ref{tab:params_values}. {STK is a highly capable modeling software that is currently used by NASA for contact scheduling in the Near Space and Direct to Earth (NSN/DTE) networks, as well as for theoretical modeling of satellite missions.} Relays in each chain work in decode-store-and-forward mode and have the capability of hard switching between RF and FSO transmitter/receiver, making the system  suitable for hybrid solutions. AGI STK communication module provides \textit{Complex} transmitter/receiver models for RF and \textit{Laser} transmitter/receiver models for FSO communication. The variable parameter settings for both transmitter and receiver of RF and FSO components in uplink (UL) chain are given in Table~\ref{tab:params_values}. Further, some parameters have fixed values, which are not shown in this table. For instance, the RF and FSO antenna efficiencies are set to $55\%$ and $70\%$, respectively, and all FSO links support communication at $1550$ nm ($193414$ GHz). Further, the FSO receivers are set to have a noise figure of $3$~dB and noise temperature is set to be $273$~K. This table also depicts the statistical values of BER for both RF and FSO links and mean propagation distances (and delay) in each link. For the input settings of RF and FSO transceivers given in Table~\ref{tab:params_values}, we depict the mean propagation loss and mean SNR for RF and FSO links in Fig.~\ref{fig:prop_loss_snr}. The first column of bar plots on $y$ axis (going front to back) describes when all three hops are FSO links (shown by \textit{OP-OP-OP} on right side), while the second column lists all RF links. These figures show only one case of hybrid links, where the proximity links are RF while the trunk link is optical (shown by RF-OP-RF). Hence, this column selects values from the previous two chains accordingly. These results can also be further extrapolated to different combinations of hard-switched hybridized links by selecting the appropriate bars from the all RF and FSO chains. Analysis of both Fig.~\ref{fig:prop_loss_snr} and Table~\ref{tab:params_values} reveals that the optical link, in comparison to RF, faces much greater propagation loss, however the much higher antenna gains compensate to achieve even greater SNR and BER values. Each row (left to right) in these plots show the each respective hop, E2L, L2G, and G2M, of the end-to-end communication chain between Earth and Lunar Facilities.

A maximum RF data rate of $15$ Mbps is achieved with the typically expected BERs and the antenna diameters given in the table, transmit powers of $1$ dBW, and an operating frequency of $10$ GHz in the Earth and Lunar proximity links. Comparing this performance for FSO links with same transmit power, there is significant increase in the data rates with much lower BER values, due to higher operating frequency, and higher antenna gains, which yield more gains in link budget analysis. Hence, the FSO links can clearly support much higher data rates. Similarly, FSO performance is superior for the trunk link between SmallSats and GW, even when RF frequency is increased to $34$ GHz and larger antenna diameters are employed, sufficient to support $1$ Mbps over the long distance. By comparison, for this link, the FSO can support about $300$ Mbps with the acceptable BER values as shown in table. The superior performance of FSO is achieved assuming the error free pointing and alignment between the transceivers. {It is also observed that the FSO links depict better BER than the RF counterparts because of the inherent higher operating frequencies, narrower antenna aperture, and higher gains in the FSO systems. Although the losses are higher in FSO case as depicted in Fig.}~\ref{fig:prop_loss_snr}, {the SNR remains higher than the RF counterpart resulting in improved BER values.} 

These results for different RF and FSO links can be joined with the access results to extrapolate the maximum data which can be transmitted for the Earth to Lunar facility. Specifically, the maximum data rates for each hop in both RF and FSO links are given in Table~\ref{tab:params_values}. Assuming the \textit{RF-OP-RF} communication chain with \textit{4-by-4} constellation of SmallSats, we have observed the chain access of $41564$ seconds. The overall per day data sending capacity with this kind of chain is calculated as, $41563(1)/8=5.195$~GB for each proximity link and $41564(300)/8=1558.650$~GB for the trunk link. The results of the average propagation distances and delays for both proximity and trunk links are also given in the last two columns of the Table~\ref{tab:params_values}. These results show that the trunk link has the highest of delays among all links, whereas the first hop link has the lowest delay.

\section{Conclusions, Challenges, and Future Research Directions}
\subsection{Architectural and Design Challenges}
This study provides an analysis of hybrid RF/Optical links for Earth-Moon communication, its performance evaluation, and access time in case of  different SmallSat constellations. Although these results are based on theoretical models, and mostly cover best case scenarios (i.e., upper limits), they can provide valuable guidance for designing a lunar communication architecture. For example, two of the key performance indicators (KPIs) in a communication system are data rate and BER. This study shows how these KPIs can be achieved and how they depend on the signal quality (SNR), modulation schemes, and hardware design parameters such as antenna gains, and operating frequencies.

Designing hybrid RF/FSO systems for space communication is not trivial, and requires tackling many challenges from a hardware integration point of view to network design and optimization. Although this work provides a basic visibility analysis on the various constellation, and link level evaluation of communication parameters, a holistic system level analysis is required with large number of earth and lunar facilities, flexibility of satellite constellation sizes in various orbits and the option of utilizing the Lunar GW. The choice of critical design parameters of Physical, MAC (medium access control), and Network Layers and their impact on the system level performance to achieve the high data rates and reliability, continuous coverage and minimum latency is required to be explored. This analysis with various possibilities of satellite constellations in tandem with PAT errors on individual RF/FSO and hybrid RF/FSO links is also an important research direction.

Effective pointing and acquisition techniques are also obvious problems to explore for the establishment of optical connections between the SmallSats and GW, especially with larger numbers of satellites. We aim not to completely turn off the RF link, but utilize it as a low data rate connection for sharing the feedback control messages. We have assumed a single Earth facility as the communication source, however, the network level analysis of multiple access schemes, interference from different satellites, and impact on the supportable data rates are yet to be explored for lunar communication systems. Another possible research direction is to utilize the soft-switching-based hybrid system where adaptive channel coding is assumed to support both types of links. In a soft-switching-based method, channel coding methods can overcome the disadvantage of hard-switching by coordinating the data transmission in both links using channel coding methods, such as LDPC and Raptor codes\cite{2009_Hard_Soft}. Further, the impact of different carrier specific channel impairments is not fully explored for the lunar communication system, which could be an important area to explore, especially for designing an adaptive system which can take channel conditions as feedback parameters in deciding to switch between RF or FSO communication. A simple hybrid network could be built by designing a system with both RF and FSO transceivers mounted on the same device. One of the design challenges is to investigate the sharing of RF and optical back-end components (for baseband processing), i.e., using one or more shared processing blocks thereby significantly reducing chip space and power requirements.
\vspace{-1mm}
\subsection{Interoperability Challenges}
{Artemis 3 will be a collaboration between NASA, commercial, and international partners to establish a sustainable lunar exploration. There are research challenges to be explored in this architecture. For instance, how do we ensure that lunar surface science support systems, which include lunar terrain vehicles, habitation/mobility platforms, and in-situ science instruments, incorporate this hybrid RF/Optical scheme in their design and operational dependencies? These are the primary users of the Lunar communication architecture and there may be some risks in selecting the communication or navigation solutions prematurely. Additional questions include what are the guiding requirements for the transmitter and receiver options that will be compatible with the Hybrid RF/Optical lunar Communication Architecture to ensure E2E compatibility and interoperability between lunar proximity links, Lunar GW, visiting spacecraft, Lunar Systems and Earth, and industry/international partner assets? How is the Hybrid RF/Optical Architecture going to handle navigation and timing?}
\section{Acknowledgment}
This work is supported by the National Aeronautics and Space Administration under Grant No. 80NSSC20M0214 issued through the Oklahoma NASA EPSCoR.

\begin{IEEEbiography}[{\includegraphics[width=1in,height=1in]{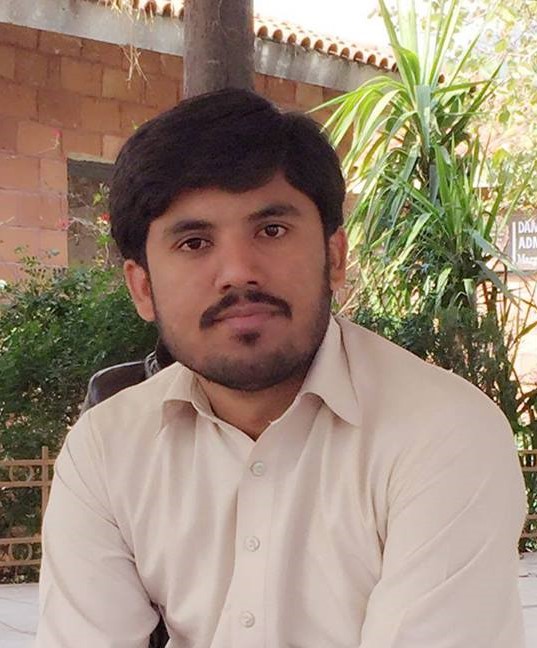}}]{Waseem Raza}
	received the B.Sc. and M.Sc. degrees in Telecommunication Engineering from the University of Engineering and Technology at Taxila, Taxila, in 2014 and 2016, respectively. He is currently pursuing towards his Ph.D. in Electrical and Computer Engineering from the University of Oklahoma, USA. His research interests include AI and ML techniques for 5G and beyond wireless networks.
\end{IEEEbiography}
\vspace*{-2cm}
\begin{IEEEbiography}[{\includegraphics[width=1in,height=1in]{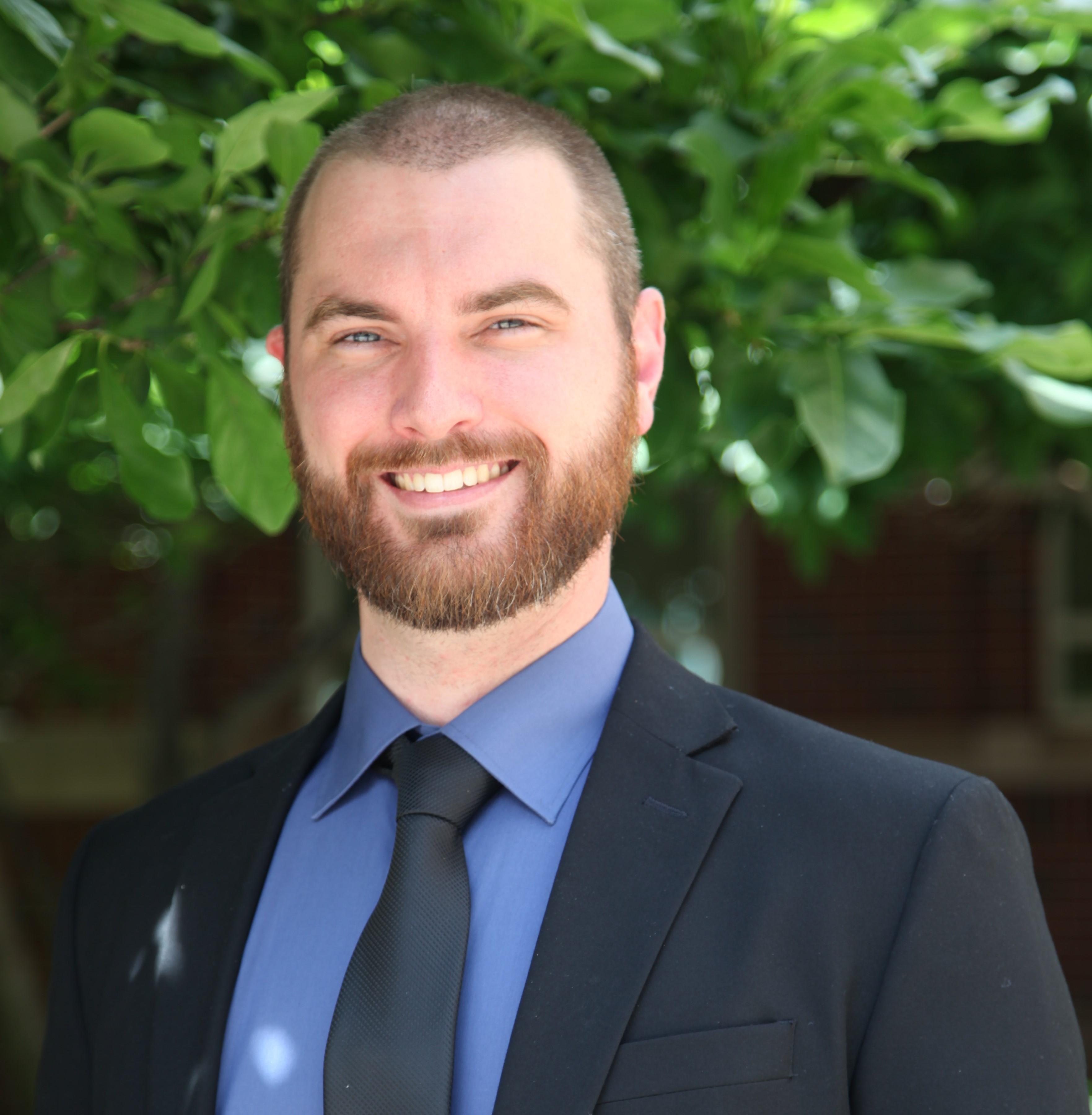}}]{Ethan Abele}
	received his master’s degree in electrical engineering from Oklahoma State University in 2016. He is currently pursuing a Ph.D. as part of the Ultrafast Terahertz Optoelectronic Laboratory (UTOL) at Oklahoma State University. His research interests include free space optical (FSO) and terahertz wireless communication technologies. 
\end{IEEEbiography}
\vspace*{-2cm}
\begin{IEEEbiography}[{\includegraphics[width=1in,height=1in]{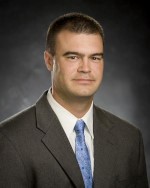}}]{John O'Hara}
	(M’05 - SM’19) received his BSEE degree from the University of Michigan in 1998 and his Ph.D. from Oklahoma State University (OSU) in 2003. He was a Director of Central Intelligence Postdoctoral Fellow at Los Alamos National Laboratory (LANL) and worked on numerous metamaterial projects involving dynamic control over chirality, resonance frequency, polarization, and modulation of terahertz waves. In 2017 he joined OSU as an assistant professor in Electrical \& Computer Engineering. His current research involves terahertz wireless communications, terahertz sensing and imaging with metamaterials, IoT, and light-based sensing and communications. He has 3 patents and around 100 publications in journals and conference proceedings.
\end{IEEEbiography}
\vspace*{-1cm}
\begin{IEEEbiography}[{\includegraphics[width=1in,height=1in]{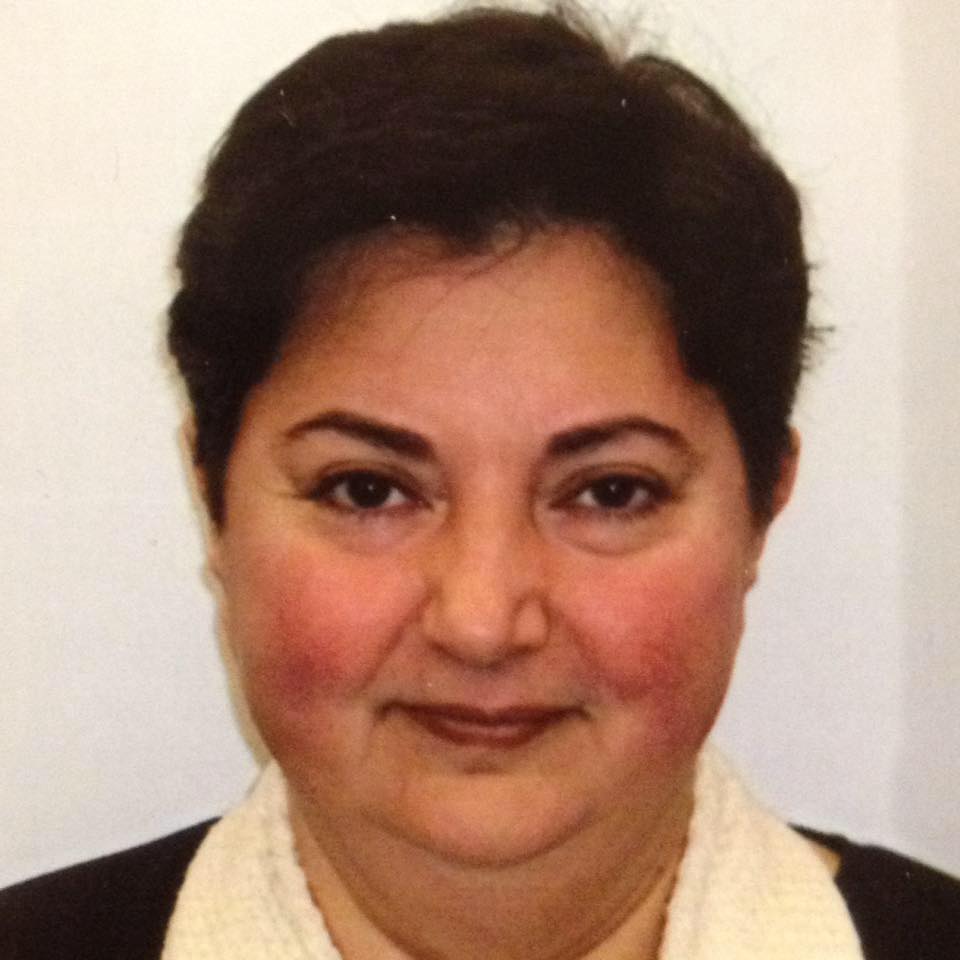}}]{Behnaz Sadr}
	received her bachelor’s degree in electrical and electronics engineering from Bilkent University, Ankara, Turkey. She graduated from Oklahoma State University in 2011 with her master’s degree in electrical engineering. Since 2018, she has been working towards a PhD in computer engineering at the University of Tulsa. Her current area of interest is free-space optical communication.	
\end{IEEEbiography}
\vspace*{-2cm}
\begin{IEEEbiography}[{\includegraphics[width=1in,height=1in]{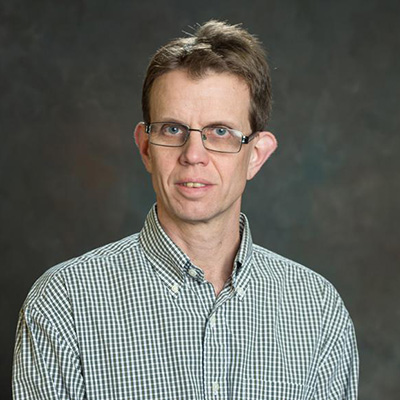}}]{Peter LoPresti}
	received his bachelor’s degree in electrical engineering from the University of Delaware in 1988 and his Ph.D. in electrical engineering from Penn State University in 1994. He has been a member of the electrical engineering faculty at the University of Tulsa since 1994, performing research in nonlinear optics, optical sensors, artificial vision, free-space optics and optical networking. He is Director of the Optical Networking Laboratory and the Integrated Spectral Analysis Instrument at The University of Tulsa. He has won numerous awards for teaching and research at the University of Tulsa, and co-edited the CRC Handbook of Neuroprosthetic Methods.
\end{IEEEbiography}
\vspace*{-2cm}
\begin{IEEEbiography}[{\includegraphics[width=1in,height=1in]{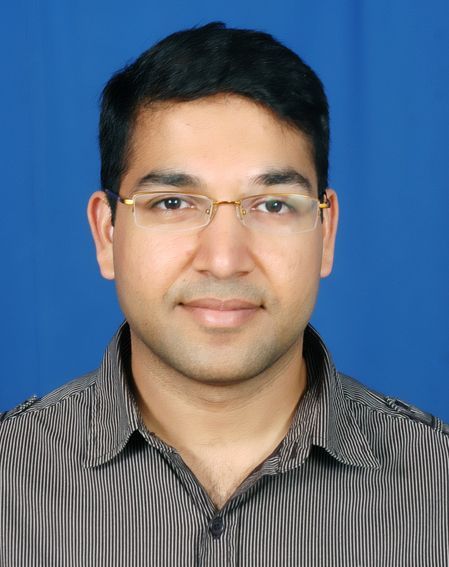}}]{Ali Imran}
	(Senior Member, IEEE) is a Williams Presidential Associate Professor in ECE and director of the AI4Networks Research Center at the University of Oklahoma. His research interests include AI and its applications in wireless networks and healthcare. His work on these topics has resulted in several patents and over 100 peer reviewed publications. He is an Associate Fellow of the Higher Education Academy, U.K.	
\end{IEEEbiography}
\begin{IEEEbiography}[{\includegraphics[width=1in,height=1in]{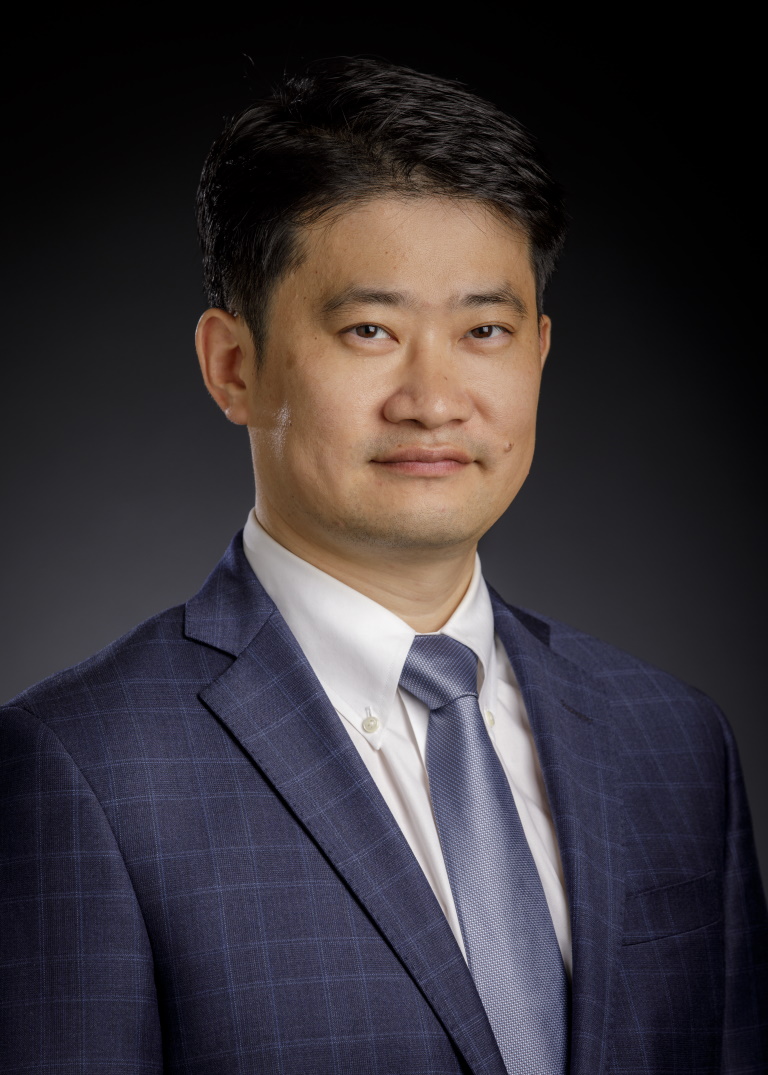}}]{Wooyeol Choi}
	(Senior Member, IEEE) received the B.S. degree in electronic engineering from Yonsei University, Seoul, South Korea, in 2001, and the M.S. and Ph.D. degrees in electrical engineering from Seoul National University, Seoul, in 2003 and 2011, respectively. From 2011 to 2018, he was with the Texas Analog Center of Excellence (TxACE), The University of Texas at Dallas, Richardson, TX, USA, first as a Research Associate and later as an Assistant Research Professor. Since 2018, he has been an Assistant Professor with the School of Electrical and Computer Engineering, Oklahoma State University, Stillwater, OK, USA. His research is focused on the design and characterization of integrated circuits and systems for various applications from RF to terahertz frequency range.
\end{IEEEbiography}
\vspace*{-2cm}
\begin{IEEEbiography}[{\includegraphics[width=1in,height=1in]{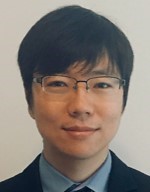}}]{Ickhyun Song}
	(S’07-M’11) received his B.S. and M.S. degrees in electrical engineering from Seoul National University, Seoul, Republic of Korea, in 2006 and 2008, respectively, and his Ph.D. degree in electrical and computer engineering from Georgia Institute of Technology, Atlanta, GA in 2016. From 2008 to 2012, he was Design Engineer at Samsung Electronics, Hwasung, Republic of Korea. In 2018, he joined the faculty at Oklahoma State University, and was Assistant Professor in the School of Electrical and Computer Engineering. From 2021, he has been with the Department of Electronic Engineering at Hanyang University, Seoul, Republic of Korea. His research interest includes extreme-environment electronics, RF/millimeter-wave circuits and systems, memory technologies, and device physics and modeling.
\end{IEEEbiography}
\vspace*{-2cm}
\begin{IEEEbiography}[{\includegraphics[width=1in,height=1in]{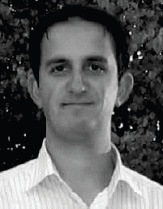}}]{Serhat Altunc}
	received his PhD in Electrical Engineering from University of New Mexico in 2007. He is currently working at Goddard Space flight Center, NASA and he is leading multiple Satellite and Ground Communication Research activities. Dr. Altunc is a flight and ground communication technologist with antenna, RF and Communication systems background. He published over 100 papers in refereed journals, book chapter, conference proceedings and presentations.
\end{IEEEbiography}
\vspace*{-2cm}
\begin{IEEEbiography}[{\includegraphics[width=1in,height=1in]{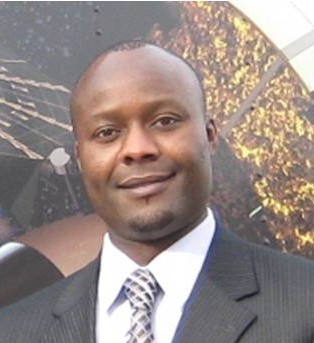}}]{Obadiah Kegege}
	serves as the Deputy Project Manager for the Geostationary Carbon Cycle Observatory (GeoCarb) mission in the Earth Science Projects Division at NASA Goddard Space Flight Center. Dr. Kegege previously served in the technical management for a number of projects: Ka-band Arraying Development; GOES Remote Ground Station development in Australia (collaboration with NOAH and USAF); and Ocean Color Instrument (OCI), part of the Plankton, Aerosol, Cloud, ocean Ecosystem (PACE) mission. Dr. Kegege had ~10 years of research and technology development supporting communication and navigation systems for space exploration. He has authored/co-authored several publications related to space communication systems and network architectures. Dr. Kegege obtained his BS in Control and Instrumentation Electronics from the University of Houston, MSEE from the University of Texas Pan American and a PhD from the University of Arkansas. Dr. Kegege is a member of IEEE, SPIE, NSBE, Instrumentation, Systems, and Automation Society (ISA), and National Technical Association (NTA).
\end{IEEEbiography}
\vspace*{-2cm}
\begin{IEEEbiography}[{\includegraphics[width=1in,height=1in]{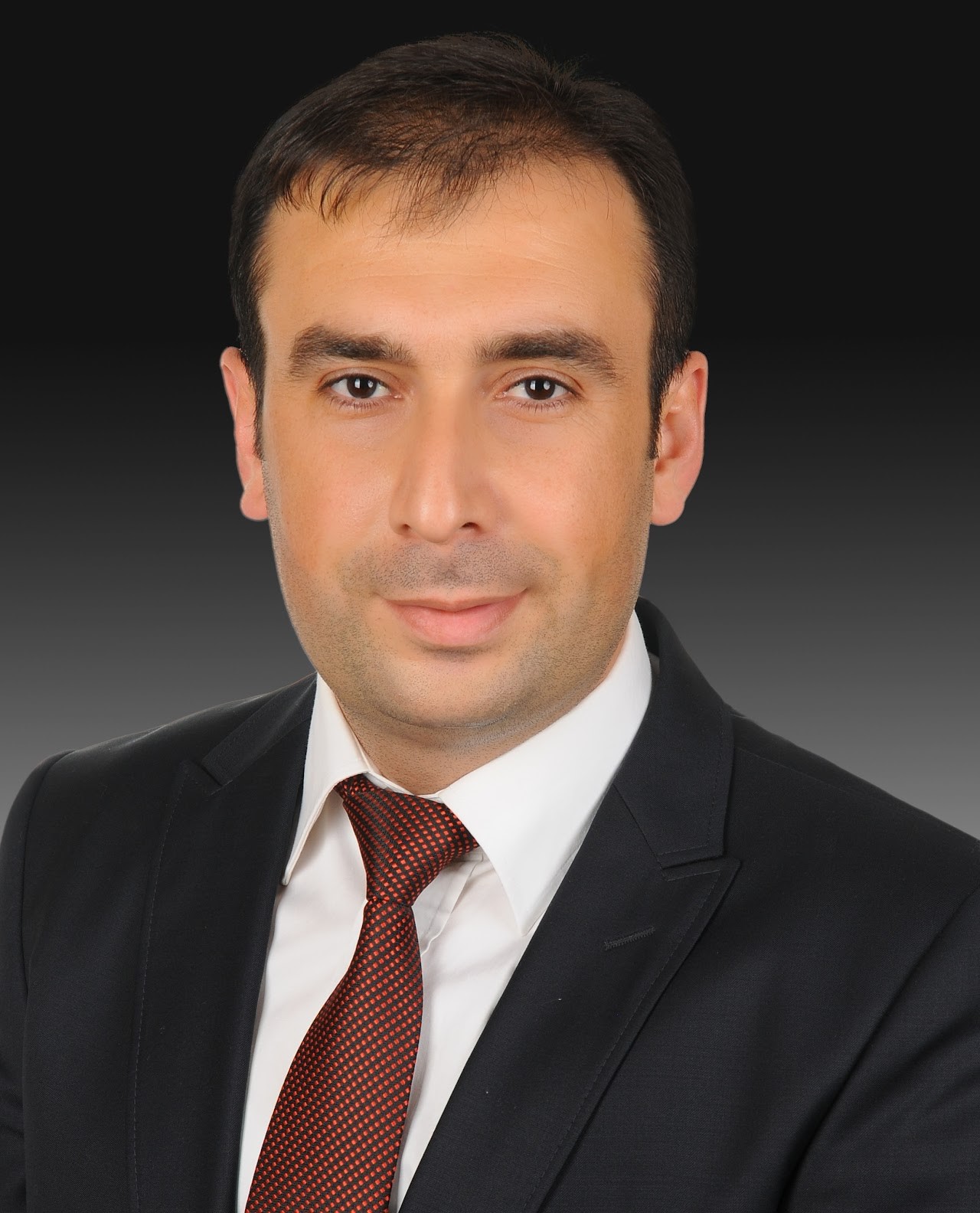}}]{Sabit Ekin}
	received his Ph.D. degree in Electrical and Computer Engineering from Texas A\&M University, College Station, TX, in 2012. He is currently a Jack H. Graham Endowed Fellow of Engineering and an Assistant Professor of Electrical and Computer Engineering at Oklahoma State University. He worked as a Wireless System Engineer at Qualcomm between 2013 and 2016.  His research interests include the design and analysis of wireless systems in both, including mmWave and terahertz communications for 5G and Beyond technologies, visible light sensing \& communications, non-contact health monitoring, and Internet of Things applications.
\end{IEEEbiography}

\end{document}